\documentclass{aa}  
\usepackage[dvips]{graphicx}
\usepackage{natbib}
\usepackage{longtable,lscape}
\usepackage{rotating}
\usepackage{txfonts}
\newcommand{\mum}{\mbox{$\rm \mu m\,$}}

\def\deg{\hbox{$^\circ$}}
\def\arcmin{\hbox{$^\prime$}}
\def\arcsec{\hbox{$^{\prime\prime}$}}

\begin{document}
\title{Molecular hydrogen jets and outflows in the Serpens south filamentary cloud} 
\subtitle{}
   \author{G. D. C. Teixeira \inst{1}, M. S. N. Kumar\inst{1}, R. Bachiller \inst{2} \and J. M. C. Grave \inst{1}
		}

   \offprints{M. S. N. Kumar; email:nanda@astro.up.pt}

   \institute{Centro de Astrof\'{i}sica da Universidade do Porto, Rua
     das Estrelas, 4150-762 Porto, Portugal 
     \and Observatorio Astron\'omico Nacional (IGN), Calle Alfonso XII, 3. 28014 Madrid, Spain\\
     \email{nanda@astro.up.pt}} \authorrunning{Teixeira et al.}
   \titlerunning{MHOs in Serpens South protocluster}
 %  \date{Received September 15, 1996; accepted March 16, 1997}
\date{\today}
% \abstract{}{2}{3}{4}{} 
% 5 {} token are mandatory
 
  \abstract
% context heading (optional)
 {} 
% aims heading (mandatory)
{To map the jets and outflows from the Serpens South star forming region 
and find an empirical relationship between the magnetic field and outflow orientation.}
% methods heading (mandatory)
{Near-infrared H$_2$ {\em v}=1-0 S(1) 2.122\mum\,-line imaging of the
  $\sim$30'-long filamentary shaped Serpens South star forming region
  was carried out. K$_s$ broadband imaging of the same region was used
  for continuum subraction.  Candidate driving sources of the mapped
  jets/outflows are identified from the list of known protostars and
  young stars in this region, which was derived from studies using
  recent {\em Spitzer}{\em} and  {\em Herschel}{\em} telescope observations.}
% results heading (mandatory)
{14 Molecular Hydrogen emission-line objects(MHOs) are identified
  using our continuum-subtracted images. They are found to constitute
  ten individual flows. Out of these, nine flows are located in the
  lower-half(southern) part of the Serpens South filament, and one
  flow is located at the northern tip of the filament. Four flows are
  driven by well-identified Class 0 protostars, while the remaining
  six flows are driven by candidate protostars mostly in the Class I
  stage, based on the {\em Spitzer}{\em} and {\em Herschel}{\em} observations.
  The orientation of the outflows is systematically perpendicular to
  the direction of the near-infrared polarization vector, recently 
  published in the literature. No
  significant correlation was observed between the orientation of the
  flows and the axis of the filamentary cloud.  }
% conclusions heading (optional), leave it empty
{}
\keywords{Stars: formation - open clusters and associations: individual (Serpens South) - 
infrared: stars - ISM: clouds - ISM: jets and outflows }

   \maketitle
%
%________________________________________________________________

\section{Introduction}

Star formation is thought to occur in molecular clouds, which are
largely known to have a filamentary structure owing primarily to the
dominant role of turbulence inside them \citep{nak11}.  Our
understanding of the details of star formation is largely benefited by
observational studies of clean templates of nearby filamentary
molecular clouds. A recently discovered filamentary cloud and its
associated protocluster \citep{gut08} in Serpens South can be used as
one such template.

The Serpens South protocluster is located in the center of a
well-defined filamentary cloud that appears as a dark patch in the
{\em Spitzer} 8$\mu$m image with an approximate extent of 30\arcmin\, in
length, and is thought to be part of the larger Serpens-Aquila rift
molecular cloud complex \citep{bontemps10,gut08,andre10}. The short
distance to this region of $\sim$260 pc \citep{gut08,bontemps10} places
it among one of the nearest star-forming regions, making it an
interesting target in the studies of star formation.

A swath of new infrared and submillimeter observations has been
recently used to study various aspects of this star-forming cloud.
\citet{gut08} used {\em Spitzer}-IRAC observations to uncover the
embedded young star population. \citet{bontemps10} presented
submillimeter observations performed with {\em Herschel} Space telescope and
identified several Class 0 protostars and candidate protostars in, and
around, the filament. \citet{sugi11} mapped the magnetic field
associated with this filamentary cloud using near-infrared
polarimetric observations. Molecular outflows in this region were
traced by \citet{nak11} using CO(J=3-2) observations. Together, these
observations, have provided a rich characterization of the Serpens
South star-forming region.

The motivation for the study presented in this paper primarily arose
from the availability of magnetic field measurements of the entire
filament. It is well known, from star formation theories, that the
bipolar outflows in young stars are ejected in the direction of the
polar axis, to which the magnetic field is also aligned
\citep{tamura}.  In particular, there is some indication that the
outflow activity is orthogonally aligned with the axis of a
filamentary cloud \citep{anath08}, owing to the nature of the
formation of prestellar cores due to the inflows along the filament
\citep{banarj}. It is possible that the presence of an ordered
magnetic field on the global scale of the filament may be related to
an improved coherence of the orientation of the outflows with respect
to the cloud axis. For example, in the DR21/W75 region, a
statistically significant coherence was found in the direction of the
flows \citep{davis07,kumar07}. The magnetic field measurements of this
region presented by \citet{kirby09} agreed quite well with the
previous result in terms of alignment between the magnetic field and
the cloud axis. However, similar outflow studies in the Orion A cloud
did not show any significant correlation of the orientation of the
outflows with the filamentary axis \citep{davis09}. This lack
of consistent results between these two different regions doesn't
allow us to settle the issue of a possible relation between the
orientation of these two components of star formation.

Although many outflow components were found for the Serpens South
filament by \citet{nak11}, the lower spatial resolution of the radio
observations, and the high density of young stars in the protocluster
makes it difficult to disentangle and relate the outflow activity with
the protostars clearly. Here, we present deep H$_2$ narrow-band
imaging study of the entire Serpens south filament to a) identify the
individual components of the flows, by building upon the results of
\citet{nak11}, b) associate the flows with the best candidate driving
source, and, c) compare the orientations of the flows with that of the
global magnetic field measured by \citet{sugi11} and also the axis of
the filamentary cloud.

\section{Observations}

Near-infrared observations were carried out on the nights of 14-17th
July 2011 using the 3.5m telescope at the German-Hispanic Astronomical
Center at Calar Alto (CAHA). Two pointings were made in order to cover
the full filament \citep{gut08}.  A near-infrared wide-field camera,
Omega2000, with a 15.3\arcmin\, field of view and a plate scale of
0.45\arcsec per pixel was used.

Observations in the H$_2$ {\em v}=1-0 S(1) 2.122\mum\, narrow-band
filter were made, a nine-point jitter sequence was applied and
repeated eight times at each pointing. An exposure time of 60 sec per
jitter was fixed, resulting in an integration time of $\sim$72 mins.  To
facilitate continuum subtraction, observations in the K$_s$ broadband
filter were obtained using a twenty-point jitter pattern, rerun one
additional time at each pointing, with an exposure time of 5 sec, so
the final integration time was $\sim$4 mins.

Each set of jittered images were median combined to obtain
a sky-frame that was then subtracted from each individual
image. Alignment of each image was performed using CCDPACK routines in
the Starlink\footnote{http://starlink.jach.hawaii.edu} suite of
software. The final mosaics were achieved using KAPPA and CCDPACK
routines in the Starlink.

The K$_s$ broadband images were psf matched and flux scaled to
perform continuum subtraction from the H$_2$ images.

\begin{figure*}
\centering

      \includegraphics[width=0.99\textwidth, angle=0]{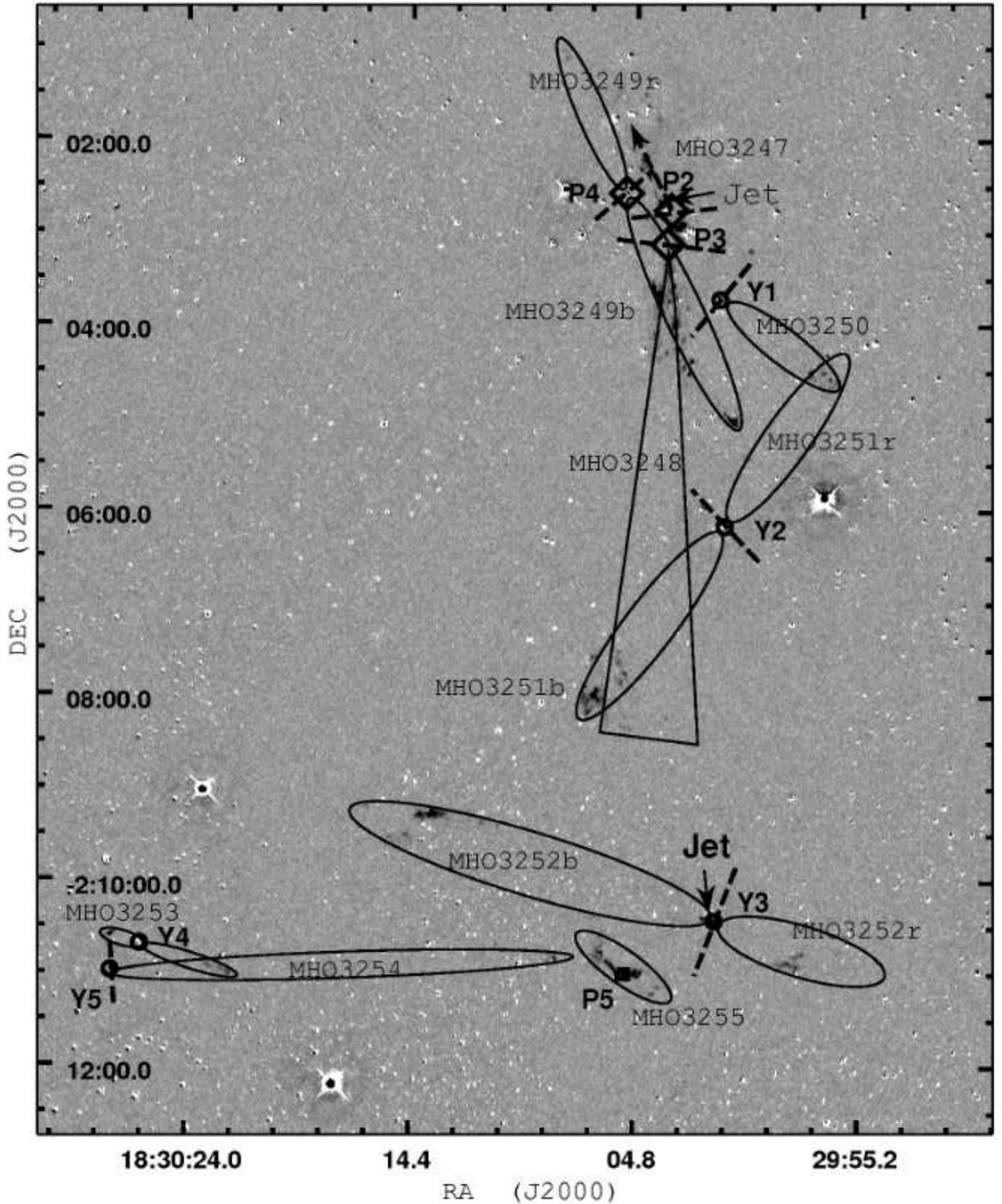}
      \caption{ Continuum-subtracted H$_2$ image of the southern part
        of the observed region, displayed with logarithmic scales. The
        sources are marked according to the following: diamonds
        represent Class 0 stars from \citet{bontemps10}, circles
        are Class I from \citet{bontemps10} and the square is a
        candidate Class 0 observed only in {\em Spitzer} 24 \mum, the labels
        P2, P3, P4, P5, Y1,Y2,Y3,Y4 and Y5 are the names we used to refer to 
        each source throughtout our work. For
        more on the outflows and driving sources see Table.\,1. Also
        visible are dashed lines representing the polarization vectors \citep{sugi11}.}
      \label{fig:outsouth}

\end{figure*}

\section{Results}

The continuum-subtracted H$_2$ images from our observations cover
almost the entire Serpens south filament (roughly 30\arcmin\, in
length). However, only two sub-regions display all the detected H$_2$
emission. Figure.\, 1 shows the southern part of the filament where
most of the H$_2$ emission is detected. The northern part of the
filament is less active, the region with detected emission is shown in
Figure.\, 2.

In these figures, the detected H$_2$ features are labeled by Molecular
Hydrogen emission-line Object (MHO) names following the catalogue
system described by \citet{davis10}.  When MHOs are associated with
clear bipolar outflows, we use the same catalogue number for both
lobes, but we add the suffix 'b' or 'r' if it is, respectively, the
blue-shifted or red-shifted lobe. 

In naming the MHO features, we first identified all possible bipolar
flows and their driving sources as explained below. We started by
identifying the presence of bowshock-like features. Then we extended a
line tracing the middle of the bowshock until we reached a possible
outflow counterpart. When this counterpart exists, the driving source
of the MHO should be along the traced line.  The driving sources were
assumed to be any of the previously identified Class 0 objects
\citep{bontemps10} or Class I objects \citep{gut08}.  In almost all
cases a suitable driving source was found. When a bipolar outflow is
identified, the MHO lobe further away in the plane of the sky from the source 
is assumed to be the blue-shifted lobe while the one closer to the source is 
assumed to be the red-shifted lobe.

After the well-defined bipolar flows were identified, the remaining
MHO features were associated with the most likely driving source, with
the aid of {\em Spitzer} data and the CO maps published by
\citet{nak11}. In Figure.\,1, all identified flows are marked by solid
lines. Diamond symbols represent Class 0 protostars and open circles
show Class I young stars. Figure.\,5 is a colour
composite made using the {\em Spitzer} MIPS 24$\mum$ image as red,
IRAC 8.0$\mum$ image as green and H$_2$ image as blue. This figure
covers exactly the same region displayed in Figure.\,1 and reveals the
dark filament. The CO flows discovered by \citet{nak11} are also
overlaid on this figure which demonstrates how well the identified
H$_2$ flows correspond to the CO emission.

A total of ten flows are identified in the entire Serpens South
filament (Figs.1 \& 2), out of which five are clearly identified as
bipolar. In the remaining, we detected H$_2$ emission only from one of
the lobes. Two well-defined jet-like features and one bipolar nebula
deleniated by a dark lane (indicating a disk) are also
identified. These interesting features are shown in Figure.\,3.

In the following we first discuss in detail each flow explaining their
morphology and the nature of the identified driving sources. Table.\,1
lists the coordinates and morphology of the MHOs. In Table.\,2 we list
the driving sources, their evolutionary class, and their visibility in
various infrared and millimeter bands.

\begin{table*}
 \centering
 \caption{List of MHOs and their sources}
 \label{tab:outflows}
\begin{tabular}{lccccc}

 \hline\hline\\
MHO & Ra & Dec & Size & Source & Comments\\

\hline

MHO2214b & 18:29:39.8 & -01:51:28 & 49.5\arcsec\ & P1 & Bipolar, bowshock\\
MHO2214r & 18:29:36.0 & -01:50:40 & 40.0\arcsec\ & P1 & Bipolar\\
MHO3247 & 18:30:04.5 & -02:02:17 & 45.5\arcsec\ & P2 & Single lobe, jet-like\\
MHO3248 & 18:30:03.4 & -02:03:48 & 317.1\arcsec\ & P3 & Single lobe, fan-shapped\\
MHO3249b & 18:30:00.6 & -02:05:02 & 104.5\arcsec\ & P4 & Bipolar, clear bowshock\\
MHO3249r & 18:30:07.8 & -02:01:06 & 169.5\arcsec\ & P4 & Bipolar, faint arc\\
MHO3250 & 18:29:56.2 & -02:04:37 & 92.3\arcsec\ & Y1 & Single lobe, bowshock\\
MHO3251b & 18:30:06.8 & -02:08:10 & 152.3\arcsec\ & Y2 & Bipolar, bowshock\\
MHO3251r & 18:29:56.8 & -02:04:33 & 129.5\arcsec\ & Y2 & Bipolar, faint arc\\
MHO3252b & 18:30:13.5 & -02:09:15 & 237.9\arcsec\ & Y3 & Bipolar, clear bowshock, jet\\
MHO3252r & 18:29:58.1 & -02:10:56 & 62.9\arcsec\ & Y3 & Bipolar, faint bowshock\\
MHO3253 & 18:30:25.8 & -02:10:43 & 25.3\arcsec\ & Y4 & Single lobe, bowshock\\
MHO3254 & 18:30:22.1 & -02:10:59 & 284.1\arcsec\ & Y4 or Y5 & Single, confusion\\
MHO3255 & 18:30:05.3 & -02:11:00 & 52.4\arcsec\ & P5 & Bipolar, embedded source\\

\hline
\end{tabular}
\end{table*}

\subsection{Catalogued MHOs} 

{\it MHO 2214b,MHO2214r:} MHO2214 is a well-defined bipolar outflow
(Figure.\,2), with collimated components.  
 Its blueshifted component had been previously discovered by \citet{con07} but our observations
were able to reveal its redshifted counterpart. This enabled us to discover its bipolar nature.
The central knot of the
blue-shifted component is a distinct bowshock. The source P1,
identified as the driving source, is classified as a protostar by
\citet{bontemps10}, and is located equidistantly from both MHO features.

{\it MHO 3247:} This MHO feature is represented in Figure.\,1 by a
dashed arrow, indicating its probable direction.  It is a jet-like
feature originating from the source P2, classified as a Class 0
protostar by \citet{bontemps10} and located in the midst of the
protocluster discovered by \citet{gut08}. No discernable MHO
counterpart is found for this feature. Note that this region is the
central part of the filament, associated with the dense cluster of
young stars and intense, multiple outflow activity \citep{nak11}
causing much confusion.\\

{\it MHO 3248:} This is a bright MHO feature with a relatively wide
opening angle. When compared with the CO maps by \citet{nak11}, it
shows clear indications of extending to the south by $\sim$317
arcseconds touching the bowshock feature identified as MHO3251b. The
most likely driving source of this MHO feature is the Class 0 object
marked as P3 in Figure.\,1, and is located in the middle of the
Serpens South protocluster. This MHO feature appears to be associated
with the blue lobes MHO3249b and MHO3251b, outflows with which it has,
likely, physically interacted.\\

{\it MHO 3249b, MHO3249r:} These MHO features are identified to
delineate a bipolar flow driven by the Class I young star P4
\citep{bontemps10} that appears as an infrared point source and is
associated with significant millimeter continuum emission. The
southern, blue lobe ends in a clear, sharply defined bowshock. The
red-shifted component has a single, poorly-defined knot but the fact
that it is well-aligned with the source and the blue-shifted lobe
allows us to determine that they are part of the same bipolar flow.
MHO3249b shows indications of crossing and/or being physically in contact with
the outflow lobe associated with MHO3248. Nevertheless, the
well-defined bowshock associated with this lobe unambiguosly
identifies the outflow lobe and the driving source.\\

{\it MHO 3250:} MHO3250 is composed of two visible H$_2$ knots and
delineates one of the lobes of an outflow driven by the Class I young
star marked as Y1. There is indication that this poorly defined MHO
crosses paths with another nearby flow associated with MHO3251r.\\

{\it MHO 3251b, MHO3251r:} These MHO features are associated with a
well-identified bipolar outflow, originating in the Class I young star
Y2 \citep{bontemps10}. The red and blue lobes are nearly equidistant
from the source (see Figure.\,1) suggesting that this flow is in the
plane of the sky. This fact is further supported by the detection of a
clear, bipolar-shaped nebula visible in the K$_s$ band images. The
bipolar nebula is divided by a dark strip showing a clear sign of the
presence of an edge-on disk in the source. This edge-on disk is zoomed
and displayed in Figure.\,3a. The MHO features align perpendicular to
the disk feature, the blue lobe MHO3251b terminates in a well-defined
bowshock and MHO3251r is composed of two knots. The presence of
another MHO, MHO3250, near the red-shifted component is the source of some
confusion but not as to significantly change our conclusions.\\

{\it MHO 3252b, MHO3252r:} These MHO features are away to the south
of the confusing central protocluster. MHO3252b consists of a
clear H$_2$ jet originating from the Class I young star marked as Y3
\citep{bontemps10}. The YSO is visible even in the near-infrared K$_s$
band and is also associated with a clear millimeter core identified
from {\em Herschel} observations. The blue lobe driven by this source,
terminates in a distinct arc-like H$_2$ feature, while the red-shifted
lobe is traced by a group of H$_2$ knots.  This flow is bright enough
to be visible prominently even in the color composite image (see
Figure.\,5).\\

\begin{table}
 \centering
 \caption{ List of driving sources}
  \label{tab:sources}
\begin{tabular}{lccccc}

  \hline\hline\\
Source & Ra & Dec & Source Type & Wavelengths(\mum) \\

\hline
P1 & 18:29:38.0 & -01:51:01 & Class 0 & 24 \\
P2 & 18:30:03.2 & -02:02:45 & Class 0 & 8-24\\
P3 & 18:30:03.6 & -02:03:11 & Class 0 & 24 \\
P4 & 18:30:05.2 & -02:02:34 & Class I & 3.6-24 \\
Y1 & 18:30:01.2 & -02:03:43 & Class I & 2.15-24\\
Y2 & 18:30:00.9 & -02:06:10 & Class I & 2.15-24\\
Y3 & 18:30:01.3 & -02:10:26 & Class I & 2.15-24\\
Y4 & 18:30:25.8 & -02:10:43 & Class I & 2.15-24\\
Y5 & 18:30:27.2 & -02:11:00 & Class I & 2.15-24\\
P5 & 18:30:05.3 & -02:11:00 &  & 24 \\

\hline
\end{tabular}
\end{table}

{\it MHO 3253:} MHO3253 could be classified as a bipolar 
outflow but since part of the outflow crosses MHO3254 we only address 
the northern lobe, a clear
arc-like structure originating in the nearby Class I source, Y4
\citep{bontemps10}.\\

{\it MHO 3254:} This feature is probably associated with a flow driven
by a Class I YSO, marked as Y5 in Fig.\,1. The eastern part of this
ouflow shows a single knot in the direction of the source which
overlaps with the possible bipolar flow originating from MHO3253,
causing some confusion.\\

{\it MHO 3255:} This H$_2$ feature has the morphology of a wiggly
bowshock. At the outset, it is likely the tip of a very long flow
originating from the Class 0 source P3, that is situated at the center
of the protocluster. A careful analysis of {\em Spitzer}
MIPS-24\mum\ images, nevertheless, reveals a faint source marked P5 in
Figure.\,1, located in the middle of this H$_2$
emission. Furthermore, this faint MIPS source is also associated with
weak millimeter continuum emission identified by {\em Herschel} observations as
candidate protostar. Therefore, we suggest that MHO3255 is a weak
outflow with a very young age driven by the candidate protostar
P5. The wiggly bow-shape of the H$_2$ features may then be due to the
motion of this source in the plane of the sky.  \\

Finally, we examined the {\em Spitzer} 4.5\mum images analized by
  \citet{gut08} to uncover any hidden census of outflows that might
  have escaped detection in our near-infrared data.  The {\em
    Spitzer}-IRAC 4.5\mum band is known to encompass a bright H$_2$
  emission line, and, the extinction at this wavelength is roughly
  half as much as in the K band \citep{flah07}. Although almost all of
  our detected outflows are also present on those images there is no
  visible indication for the existence of additional outflows in the
  studied area. Indeed, the H$_2$ data presented here provide the
  certainty of tracing pure emission line features unlike the 4.5\mum
  data where judgement is based on comparision with other bands and
  not real continuum subtraction. Further, our H$_2$ data offers a
  superior spatial resolution disentangling the morphological
  details. Not detecting additional features from {\em Spitzer} data
  is an indication that the outflow sample traced by our data is
  quite complete and the extinction in this cloud is similar to other
  nearby regions of star formation where H$_2$ narrow-band images
  effectively serve as jet/outflow tracers.

\begin{figure}
\centering
\vspace{5pt}
      \includegraphics[scale=0.3, angle=0]{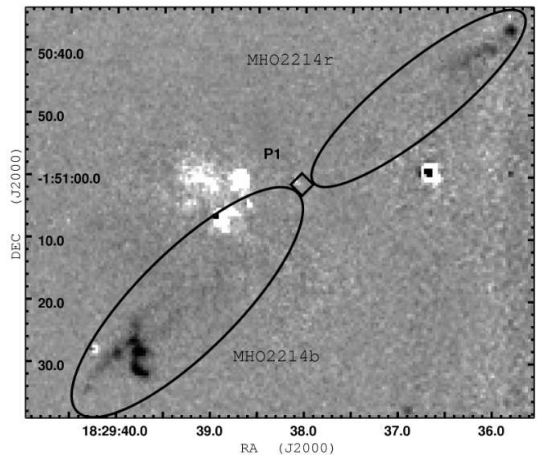}
      \caption{Continuum-subtracted H$_2$ image of the northern part of the 
      observed region, displayed with logarithmic scales. This bipolar outflow
      well aligned with Class 0 protostar P1 \citep{bontemps10}.}
      \label{fig:outnorth}

\end{figure}

\begin{figure*}
\centering

      \includegraphics[scale=0.5, angle=0]{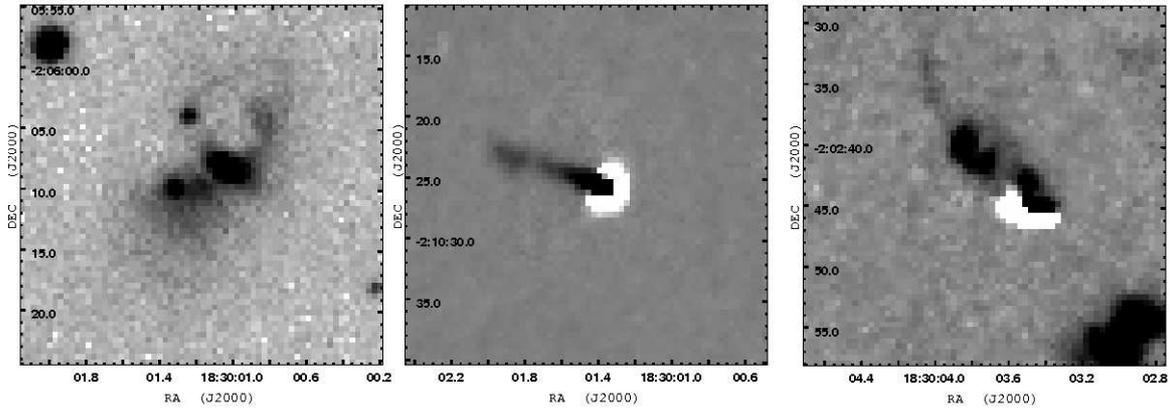}
      \caption{From left to right the images are centered on: the
        bipolar nebula of Y2 visible in the K$_s$ broadband filter,
        the outflow jet from Y3, on a continuum-subtracted image, and
        the the jet from P2, also in a continuum-subtracted image.}
      \label{fig:bipolarzoom}

\end{figure*}

\begin{figure}
\centering
     \vspace{25pt}
     \includegraphics[height=5cm, angle=0]{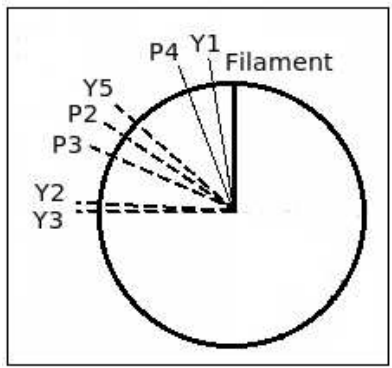}
     \caption{ Positional angles(PA) of the sources polarization relative 
      to the PA of the filament(0 degrees). Solid and dashed lines show PA$<$45\deg\, 
      and PA$>$45\deg\, respectively.}
      \label{fig:PAngles}

\end{figure}

\section{Discussion}

In the studied region we found a total of ten outflows, five of which
are clearly bipolar and only a single lobe was detected in the
remaining five MHOs. We then studied the collimation factors of each MHO,
 defined as the ratio between the length and the
  width of each outflow.  At least two of the observed MHOs have high
  collimation factors, particularly MHO3249r and MHO3254, with
  collimation factors of 11.2 and 15.6 respectively. These examples
of extreme collimation can be directly linked to the young nature of
these sources and may be an indication of the presence of high
velocity components \citep{bach96}.  On the other hand, we also
observe some outflows that present low collimation factors (between 2
and 5), MHO3253, MHO3252r and MHO2214b with factors of 2.5, 4.2 and
4.6 respectively. These low collimation factors point towards the fact
that the origin of these outflows lies in ambient gas picked up by
stellar winds \citep{bach96}.

We also performed a rough comparison between the orientation of
the MHOs and the axis of the cloud filament. Outflows were considered
parallel to the filament if their orientation angle with respect to
the axis of the filament is less than 45$\deg$ and perpendicular when
the angle was greater than 45$\deg$. This comparison showed that
$\sim$71\% of the flows are perpendicular and $\sim$29\% are parallel
to the filament. In Figure 4 we plotted the positional
angles(PA) of the known polarization vectors relative to the PA of the filament 
or subfilament where each source is located, in order to observe the 
distribution of parallel and perpendicular vectors. 
Although these fractions are similar to the values
obtained by \citet{anath08} ($\sim$72\% perpendicular and $\sim$28\%
parallel), we stress that our sample is not large enough to be
statistically significant. Although we cannot imply the existence 
of a coherent pattern in the orientation of the outflows due to 
poor statistics, our analysis demonstrates that the fraction of 
flows perpendicular to the filament are clearly larger than those 
parallel to the filament.

Nevertheless, we find a better correlation between the orientation of
the outflows with respect to the local magnetic field. To obtain this
result, we compared the direction of MHO flows with that of the
near-infrared polarization vectors of the driving sources
\citep[obtained from the data of][]{sugi11}. We were able to find
polarization vectors associated with seven identified driving sources,
namely P2, P3, P4, Y1, Y2, Y3 and Y5. These vectors are shown as
dashed lines in Figure.\,1. In all these cases, the outflows are
generally found to be perpendicular to the polarization vectors. In
particular, the polarization vector of source Y2, traces perfectly the
disk of that young star (see Figure.\,3) which is identified as a dark
lane cutting across a bipolar reflection nebula, and is found to be
perpendicular to MHO3251.

  Polarization measurements of light from young stars are known to
  represent the disks and envelopes rather than the embedded
  proto-star. This is because, the light originating in the young star
  is reprocessed in the dense envelopes surrounding them before
  reaching the observer. Therefore, the observed polarization vectors
  are usually found to be parallel to the disk axis which is
  perpendicular to the axis of the stellar magnetic field
  \citep{tamura,pereyra,murakawa}. In fact, the level of polarization
  is found to increase in younger sources and/or edge-on disks
  \citep{greaves97}.

It is known that outflows follow the open magnetic lines of the
source, which means that their origin should be near the magnetic
poles of the source \citep{kon}.  Since the majority of the observed
outflows are perpendicular to the known polarization vectors we can
conclude that the observed polarization vectors are perpendicular to
the magnetic field of the driving sources. Given this close relation
between the local magnetic field and the arrangement of MHOs, it is
clear that the outflows are aligned with the local magnetic field.
While the large scale orientation of the polarization vectors roughly
run perpendicular to the axis of the filamentary cloud, local twists
are found to exist from the observations of \citet{sugi11}. The
location of the driving sources along such twists may, therefore,
indicate the position of star forming cores. As accretion proceeds,
the magnetic field appears to become distorted, so we don't expect or
observe any coherent alignment between the global magnetic field and
the orientation of the MHOs \citep{pereyra}. This is likely one of the
reasons we do not find a high degree of coherence between the
orientation of the flows with the filamentary axis. Other factors such
as turbulence, the morphology of the filament at smaller scales and
the distribution of cores can also strongly influence the deviations
of the local magnetic fields from the global field.

%\onlfig{5}{ 
  \begin{figure*}
  \centering
  \includegraphics[width=0.75\textwidth, angle=0]{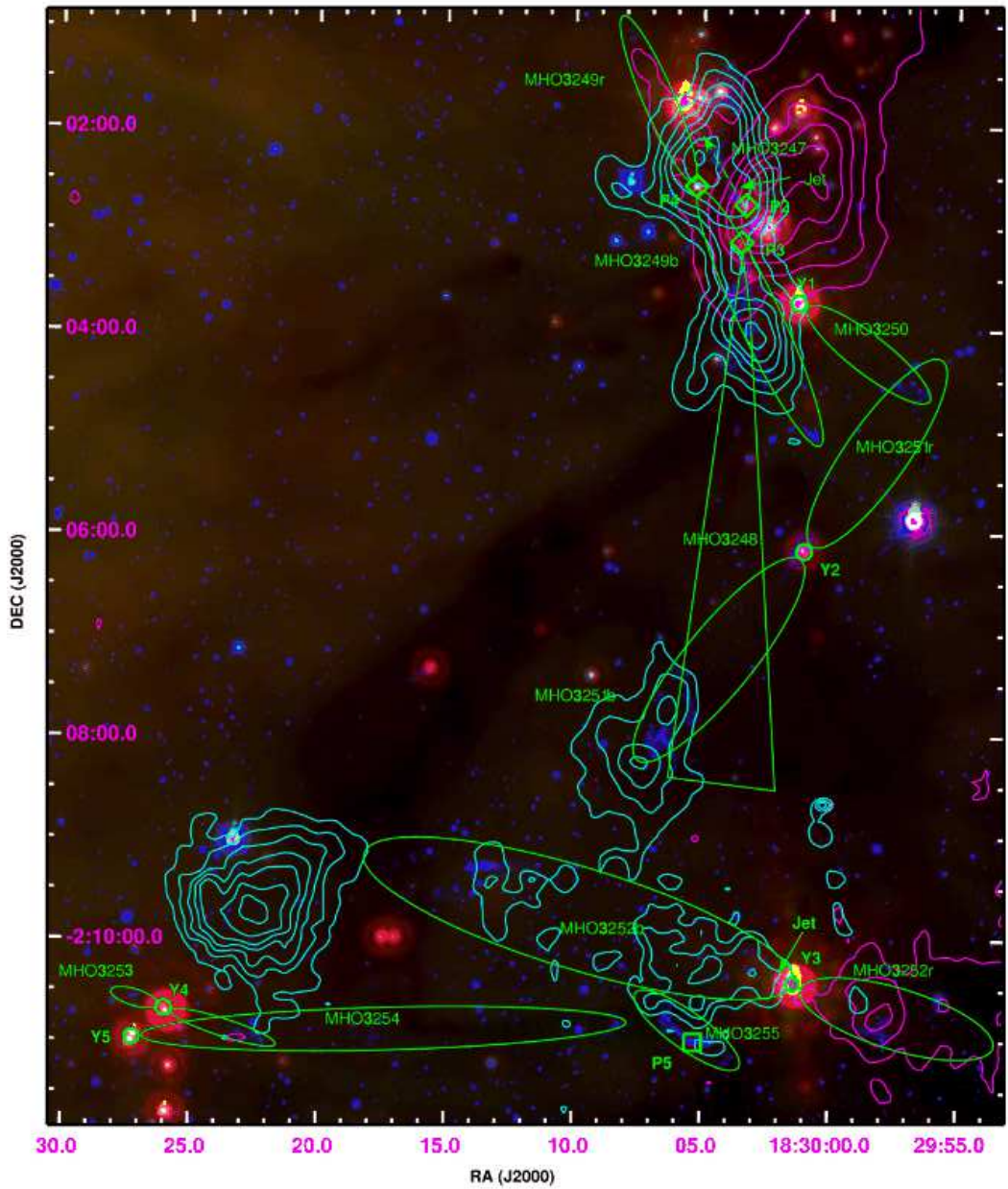}
  \caption{A colour image of the southern region studied.  {\em
      Spitzer} MIPS 24\mum\, is red, {\em Spitzer} MIPS 4.5\mum\, is
    green and H$_2$ narrowband image as blue. The images use a
    logarithmic scale. The red and blue shifted CO emission mapped by
    \citet{nak11} is displayed using, respectively, pink and blue contours. }
  \end{figure*}%}

\section{Summary and Conclusions}

We discovered new MHO features in the filamentary molecular cloud
''Serpens South'' associated with a protocluster, in the Aquila
Rift. We have catalogued MHO features associated with 10 outflows,
five of which are clearly identified as bipolar outflows
(Table.\,1). A comparison between our survey and previous 
images obtained with {\em Spitzer}
showed no additional MHOs in the region.
Two jet-like features, launched from the driving YSOs are
clearly identified. One edge-on disk in this region is also identified
from this study(see the bipolar nebula in Figure.\,3). Half of the detected flows are identified to be
driven by previously classified Class 0 objects and the remaining half
are driven by Class I type sources. The MHO features found here are
well associated with the CO outflows presented by \citet{nak11}. In
particular, MHO3247, MHO3248, MHO3251 and MHO3252 appear to trace the
CO maps extremely well matching the blue or red shifted lobes.

We find that two of the observed MHOs have high collimation factors,
indicating that we are dealing with high velocity flows.

Our H$_2$ imaging study has allowed disentangling and clarifying the
association between outflows and driving sources in the central
protocluster. The high concentration of outflows from the protocluster
located at the center of the filamentary cloud displayed some
confusion that did not permit an unambigous association of the
outflows and their sources. Our results also indicate that the flows
launched from the central protocluster may have interacted with each
other physically since they are launched from a small region. MHO3248
and MHO3249b are two such signatures.

We show that the near-infrared polarization vectors mapped by
\citet{sugi11} are perpendicular to the outflows and therefore to the
magnetic field of the sources. While the outflows are perpendicular to
the local polarization vectors, no specific coherence is found with
respect to the global pattern. Despite dealing with a sample that is
statistically not significant, we find that $\sim$71\% of the flows in
this region are perpendicular to the axis of the filamentary cloud, see Fig.\, 4,
an estimate that agrees well with the previous work by \citet{anath08}.

\begin{acknowledgements}
 We thank F. Nakamura for providing the CO data for overplotting in
 the colour Figure. Teixeira acknowledges support from a Marie-Curie
 IRSES grant (230843) under the auspices of which, part of this work was
 carried out. Kumar is supported by a Ci\^encia 2007 contract, funded
 by FCT/MCTES (Portugal) and POPH/FSE (EC). We also thank the staff at
 the Calar Alto observatory for their support.

\end{acknowledgements}

%__________________________________________________________________

\bibliographystyle{aa}
\bibliography{arxiv}

\end{document}